# On the Performance of Mobile Visible Light Communications

Yang Hong and Lian-Kuan Chen

Dept. of Information Engineering, The Chinese University of Hong Kong, yanghong@ie.cuhk.edu.hk

**Abstract** We experimentally characterize the performance of mobile VLC and propose using OCT precoding to combat mobility-induced performance degradation. Results show that for approximate 300-Mb/s mobile VLC transmission, OCT precoding outperforms adaptive-loaded DMT and offers significant packet loss rate reduction.

**Introduction**

Compared with the conventional radio frequency based wireless communications, visible light communications (VLC) provides the advantages of free license, enhanced physical security, and immunity to electro-magnetic interference[1]. Therefore, VLC has attracted extensive interests from both academia and industry in recent years. By utilizing schemes like multiple input and multiple output (MIMO)[2] and discrete multi-tone (DMT) with adaptive bit and power loading[3], high speed VLC transmissions (~ gigabit/s) have been reported. Recently, a blue laser, together with a phosphorous diffuser, has emerged as a promising candidate for high speed VLC-based wireless access[4]. However, most of the prior experimental demonstrations are based on stationary point-to-point (P2P) setups, i.e., LED transmitter and receiver are fixed and convex lenses are utilized at both sides to maximize the detected power, thus achieving higher data rate. For emerging applications like inter-vehicle and undersea VLC, it is essential for the system to support user terminal mobility. The system, termed as mobi-VLC, also provides enhanced flexibility and robustness for indoor applications to mobile users. The performance of a color-clustered VLC network with user mobility is numerically studied, but without experimental verification[5]. A self-adaptive space-time block coding (STBC) scheme is proposed to mitigate the interference between two VLC nodes, thus improving handover capability for VLC-based networks[6]. The experiments are conducted at several stationary points without user mobility. User mobility, in general, has significant impact on VLC systems because of the mobility-induced intensity fluctuation. Furthermore, compared with the stationary cases, inter-symbol interference (ISI) is more severe in mobi-VLC when there is multipath effect.

In this paper, to our best knowledge, we first present the experimental characterization of the impact of user mobility on the transmission performance of the mobi-VLC system. A mobile platform with controllable lateral distance and speed is implemented to emulate various user mobility scenarios within the coverage of the system. In addition, we propose the use of our recently demonstrated channel-independent OCT precoding[7] to combat mobility-induced performance degradation and compare the results with that of the conventional orthogonal frequency division multiplexing (OFDM) and the DMT with adaptive loading. Experimental results show that for the mobi-VLC transmission at a data rate around 300-Mb/s, both adaptive-loaded DMT and OCT precoding can reduce the packet loss rate of the system. Besides, the improvement of OCT precoding is more significant due to its channel-independent property.

**Experimental setup**

The experimental setup of the conventional OFDM-based VLC system with stationary transmitter and mobile receiver is shown in Fig. 1. After serial to parallel (S/P) conversion and mapping, Hermitian symmetry operation was needed before inverse fast Fourier transform (IFFT) to enable the real-valued OFDM signal. Subsequently, cyclic prefix (CP) insertion and parallel to serial (P/S) were performed and the resulting signal was fed into arbitrary waveform generator (AWG, *Tektronix* 7122C) to generate continuous time-domain OFDM signals. The output of AWG was then amplified by a tunable electrical amplifier (EA). The amplified signal, coupled with a DC bias by a bias-tee (Mini-Circuits ZFBT-6GW+), was utilized to drive a single blue laser diode (*Osram*, PL450B). The light was detected by a PIN photodiode (Hamamatsu S10784) and further amplified by a trans-impedance amplifier (TIA) circuit. The aforementioned receiver module was installed on a mobile platform with controllable lateral distance $d$ cm and speed $v$ cm/s to emulate the user mobility. A digital phosphor oscilloscope (DPO, *Tektronix* TDS7254) was utilized to record the received signal for further offline signal processing.

In this work, the block FFT size of OFDM was 256, the sampling rate of AWG was 300 MS/s,

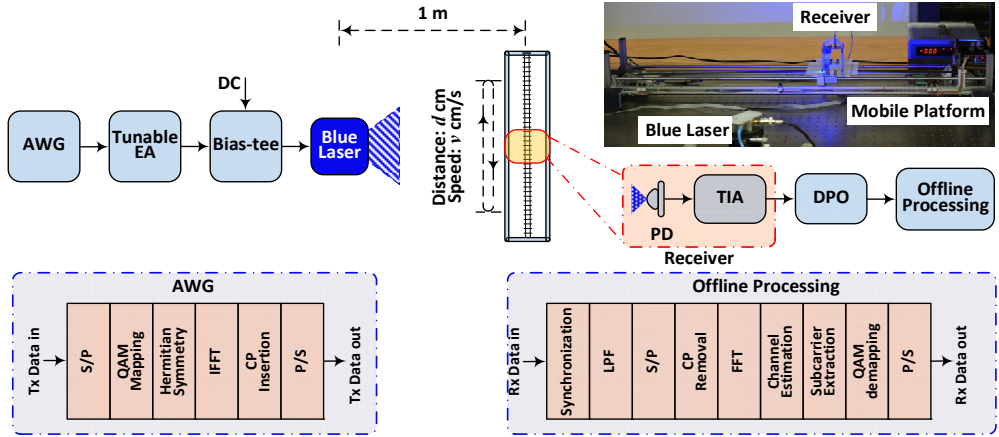

**Fig. 1:** Block diagrams of OFDM-based VLC system with user mobility. Inset: photo of the system setup.

the sampling rate of DPO was 625 MS/s, the CP length was 1/32 of one OFDM symbol and 127 out of 256 subcarriers were modulated with data in one OFDM symbol. 500 OFDM packets were captured automatically by the DPO during the movement of the receiver module to investigate the impact of user mobility and location on the system performance. Each OFDM packet consists of 200 OFDM symbols and 20 training symbols, and all the experiments were conducted under normal illumination (~400 lux).

**Results and discussions**
Considering the influence of laser nonlinearity on OFDM signals, firstly we investigate the system performance with different levels of bias voltage and driving signal intensity. Fig. 2 shows the BER performance of the system using conventional 4QAM-OFDM. Note that in the experiments, the receiver was kept stationary in the middle of the track while investigating optimal bias voltage and amplification level.

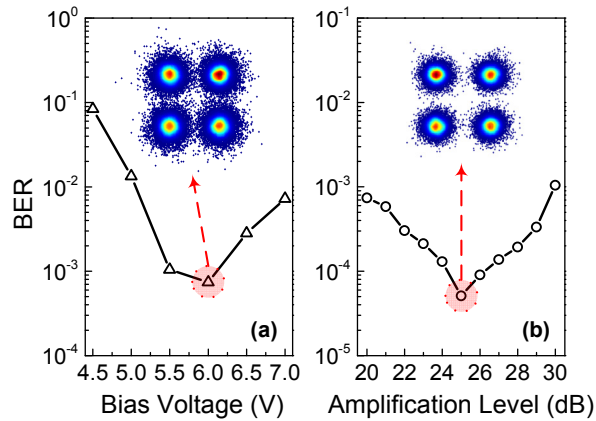

**Fig. 2:** Measured BERs versus (a) bias voltage (Amplification level is 20 dB) and (b) Amplification level of driving signal (bias voltage is 6.0 V).

As shown in Fig. 2, the optimal bias voltage for the blue laser is 6.0 V. When amplification level is 20 dB, the BER performance of the system with 4QAM-OFDM, i.e., around 300 Mb/s transmission, is $7.40 \times 10^{-4}$. With the optimal bias voltage, we increase the amplification level to investigate the optimal driving signal intensity for the laser. From the results shown in Fig. 2 (b), the optimal amplification level is 25 dB, at which the minimum BER of $5.12 \times 10^{-5}$ can be achieved. Note that the optimal amplification level lies in the medium level because of the nonlinearity induced by a larger driving signal and low SNR condition resulted from a lower driving signal.

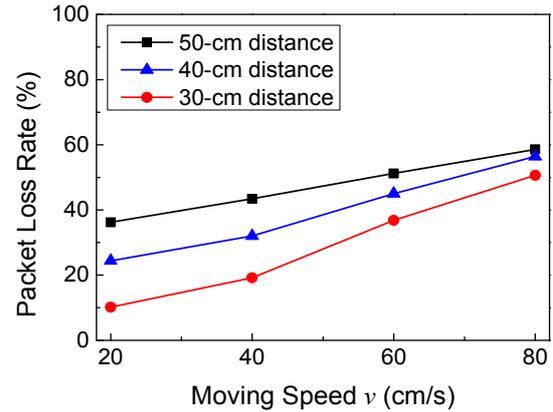

**Fig. 3:** Packet loss rate versus user mobility with different lateral distance. A lost packet means a packet with a BER larger than $3.8 \times 10^{-3}$ FEC limit.

Using the optimized bias voltage and amplification level, we further investigate the system performance under various user mobility. Fig. 3 shows the packet loss rate of the received 4QAM-OFDM packets with different moving speed while the lateral traverse distance is 30 cm, 40 cm and 50 cm, respectively. Due to the emission property of the blue laser, the packet loss rate will degrade along with the increase in the lateral distance. As shown in Fig. 3, for certain lateral distance, the performance degradation is not significant when receiver's moving speed is below 40 cm/s. However, higher moving speed will intensify the system performance degradation and the packet loss rate can be higher than 50%.

Taking 50-cm lateral distance as an example, the statistical analyses of user mobility's influence on the BER of the system are given in the form of box chart and distribution curve in Fig. 4. It is clear that with the increase in moving speed, the number of OFDM packets with poor BER performance will also increase, resulting in the ascent of the 10% to 90% distribution range, as well as the average BER of the system.

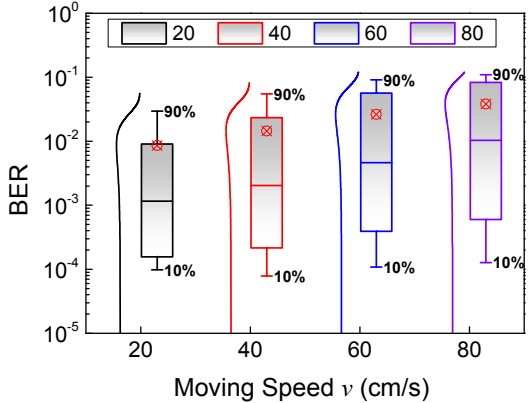

**Fig. 4:** Measured BER distributions versus user mobility with 50-cm lateral distance (Red marks denote the average BERs of the four scenarios).

To enhance the system performance, OCT precoding is utilized to combat the mobility-induced performance degradation. Meanwhile, DMT with adaptive bit and power loading scheme is used for comparison in the mobi-VLC system. For the OCT precoding, no pre-estimation is needed since it is a channel-independent scheme. For the adaptive-loaded DMT scheme, 20 OFDM packets are captured for each moving speed with certain lateral distance to estimate the channel state information, so as to adaptively allocate the bits and power to different subcarriers.

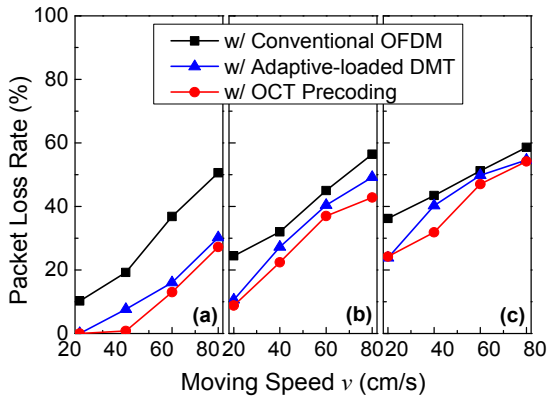

**Fig. 5:** Packet loss rate versus user mobility with lateral distance being (a) 30-cm, (b) 40-cm and (c) 50-cm.

Fig. 5 shows the packet loss rate comparison of the system with conventional OFDM, adaptive-loaded DMT and OCT precoding, respectively, under different user mobility and lateral traverse distance. It can be seen that the improvement of packet loss rate for OCT precoding outperforms that of adaptive-loaded DMT. Packet loss rates of 0, ~10% and ~20% can be achieved for 30-cm, 40-cm and 50-cm lateral distance, respectively, when the data rate of the mobi-VLC system is around 300 Mb/s and the moving speed is 20 cm/s. Besides, for all the three schemes, the packet loss rate will increase along with the increase in lateral distance of the user, as expected.

**Conclusions**

The first experimental characterization of user mobility effect on the performance of the mobi-VLC system is presented in this work. We show that the system performance will degrade with the increase in user mobility speed and the increase in lateral distance. Robustness of the OCT precoding and the DMT with adaptive loading in the mobi-VLC system are evaluated. Experimental results show that the proposed system using channel-independent OCT precoding provides better packet loss improvement than that of the adaptive-loaded DMT. Packet loss rates of 0 and ~20% can be achieved for 300-Mb/s transmission with 30-cm and 50-cm lateral distance at 20 cm/s moving speed, respectively.

**Acknowledgements**
This work was supported by HKSAR RGC grant (GRF 14200914). The authors would like to thank Mr. Alex Wai-hin Siu for preparing the mobile platform of this work.